# The intersecting brane world

B. F. Riley

AMEC NNC Limited,
601 Faraday Street, Birchwood Park,
Warrington WA3 6GN, UK

bernard.riley@amecnnc.com

**Abstract**

The scales of the Standard Model correspond to the positions of domain wall intersections on a straight line in a noncompact two-dimensional extra space. The domain walls partition an Anti-de Sitter spacetime. We show that domain wall intersections mark the orthogonal intersections of 1-cycles winding around a rectangular compact space and a hierarchy of subspaces, and that the noncompact extra space of the model is a covering space. 4-branes wrap 1-cycles of the compact spaces and intersect orthogonally in 3-branes, upon which the particles propagate. We show that particles are located precisely at domain wall intersections in the covering space.

# 1 Introduction

In a previous paper [1], we related four-dimensional mass parameters to the positions of lattice points in two infinite extra dimensions, building on the brane world models of Randall and Sundrum [2, 3]. We conjectured that the two lattices extend in two directions, y and z, in a two-dimensional extra space. Domain walls extending through the lattice points partition a bulk Anti-de Sitter (AdS) spacetime. The domain walls lie along and parallel to the sides of cells which tile the extra space. The scales of the Standard Model were shown to correspond to the positions of domain wall intersections on the line y = z. Here, we will show that domain wall intersections mark the positions of 1-cycle intersections in the covering space of a rectangular compact space.

In the Randall and Sundrum RSI model [2], the Standard Model particles and forces are confined to a 3-brane, the 'weak' brane, at a fixed point of the $S^1/Z_2$ orbifold. The Planck brane is located at the other fixed point of the orbifold. The two branes are separated by a slice of AdS spacetime, of curvature $k$. The natural scale on the weak brane is suppressed from Planck-scale by a warp factor exp(-$k$y), where y is the coordinate of the extra dimension. In the RSII model [3], there is only one brane, at y = 0 in an infinite dimension, upon which the Standard Model particles and forces are confined. Lykken and Randall [4], and Oda [5, 6], have shown that we can live on a brane, separate from the Planck brane, in the infinite dimension. Arkani-Hamed, Dimopoulos, Dvali and Kaloper have shown that gravity can be localized to the intersection of orthogonal (2+n)-branes lying in an infinite $AdS_{n+4}$ spacetime [7]. Aldazabal, Franco, Ibanez, Rabadan and Uranga have proposed a model in which chiral fermions propagate at brane intersections located at different positions in an extra space, in particular at the intersections of D4-branes wrapping 1-cycles of a 2-torus [8].

In Section 2, we will review our previous work [1]. In Section 3, we will provide evidence that the noncompact extra space of the model is the cover of a rectangular compact space. We will show that 1-cycles winding around the fundamental 1-cycles of the rectangle intersect upon the line y = z in the covering space at positions that correspond to the scales of the Standard Model. In Section 4, we will show that particles are located at the orthogonal intersections of 1-cycles on the line y = z. 4-branes wrapping the 1-cycles intersect in 3-branes, upon which the particles propagate.

## 2  Mass hierarchies from two extra dimensions

A lattice of points extending from the Planck brane with spacing d/k in a semi-infinite extra dimension corresponds, in four dimensions, to a geometric sequence of mass-energy scales that descends from the Planck Mass $M_P$ (1.221 x $10^{19}$ GeV) with common ratio exp(-d), provided that $M_P = M^*$, where $M^*$ is the Planck scale of the higher dimensional, AdS, spacetime. By computation, we found that the masses of particles of all types fall upon the levels and sublevels, described below, of two such geometric sequences, suggesting that particles are located at lattice points within two extra dimensions. Sequence 1 and Sequence 2 are of common ratio $2/\pi$ and $1/\pi$, respectively, to within 1 in $10^5$. The mass $m_i$ of the $i^{th}$ level of Sequence 1 is given by

$$m_i = (\pi/2)^{-i} M_P, \qquad (1)$$

where $i \geq 0$. The mass $m_j$ of the $j^{th}$ level of Sequence 2 is given by

$$m_j = \pi^{-j} M_P, \qquad (2)$$

where $j \geq 0$. Principal mass levels in Sequence 1 and Sequence 2 are of integer $i$ and $j$, respectively. Higher order mass levels (sublevels) are of fractional $i$ and $j$. For example, $1^{st}$ order levels in Sequence 1 are of half-integer $i$, while $2^{nd}$ order levels are of quarter-integer $i$. The dimensionless variables $i$ and $j$ also measure the corresponding distance along the two extra dimensions, from the Planck region ($i = 0$, $j = 0$), in units of $(1/k)\ln(\pi/2)$ and $(1/k)\ln\pi$, respectively. At principal lattice points in the two extra dimensions, $i$ and $j$ are of integer value. Higher order lattice points divide the space between the principal ($0^{th}$ order) lattice points into $2^n$ intervals of equal length, where n is the order.

Quark masses in the $\overline{MS}$ scheme, as evaluated by the Particle Data Group [9], are consistent with the masses of principal levels in Sequence 1. The up quark mass of the model is 3.26 MeV ($i$ = 110). The down quark mass is 8.05 MeV ($i$ = 108). The strange quark mass is 121 MeV ($i$ = 102). The charm quark mass is 1.16 GeV ($i$ = 97). The bottom quark mass is 4.48 GeV ($i$ = 94). The top quark mass is 166 GeV ($i$ = 86).

The lightest flavoured baryons are associated with principal and first order mass levels in Sequence 1. The masses of the uds baryons $\Lambda$ and $\Sigma^0$ lie either side of a principal level, while the masses of $\Lambda_c^+$ (udc) and $\Lambda_b^0$ (udb) lie close to $1^{st}$ order levels, as shown in Figure 1. The masses of the lightest flavoured mesons, $K^\pm$ and $K^0$, and the electron are situated close to principal levels in Sequence 2, as shown in Figure 1.

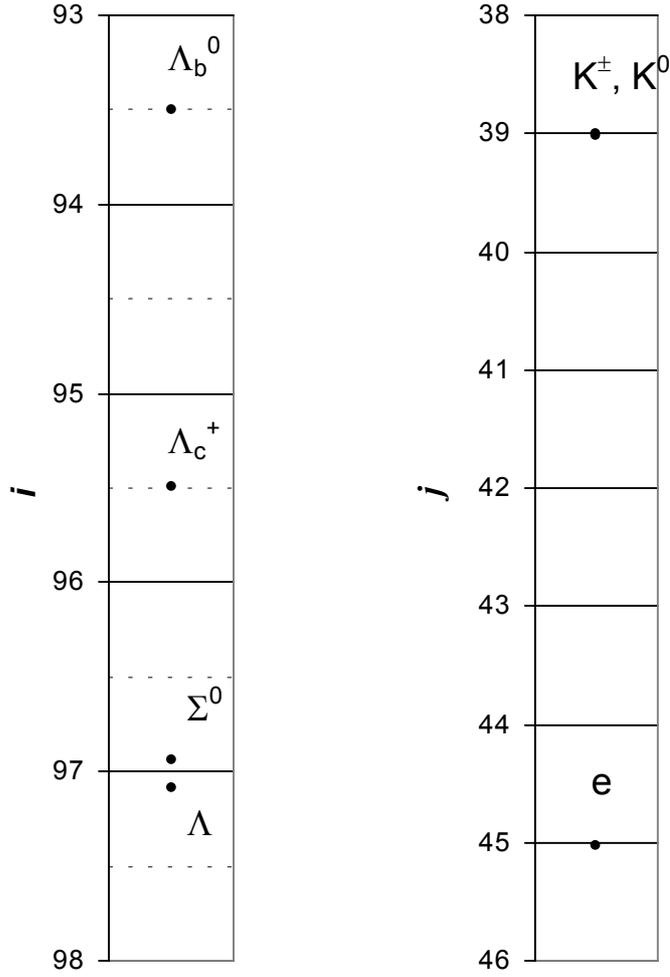

Figure 1: Values of *i* in Sequence 1 for the lightest flavoured baryons, and *j* in Sequence 2 for the K-mesons and the electron.

By conjecturing that the two lattices we have identified lie in two different directions, y and z, in a noncompact two-dimensional extra space and that domain walls extending through the lattice points partition an AdS spacetime, we have found that the major scales of physics correspond to the positions of domain wall intersections $(i, j)$ lying on the line y = z. Principal domain walls intersect at y = z where $(i/k)\ln(\pi/2) = (j/k)\ln\pi$; that is, where $i/j = \ln\pi/\ln(\pi/2) = 2.5349$. This condition is met closely at the intersection (71, 28) where $i/j = 2.5357$, at (109, 43) where $i/j = 2.5349$, and at (147, 58) where $i/j = 2.5345$. The positions of these principal domain wall intersections correspond to scales in a geometric sequence of common ratio $(\pi/2)^{38} \approx \pi^{15}$. This sequence of mass-energy levels has its own set of sublevels, also corresponding to the positions of domain wall intersections lying on the line y = z.

The position of the principal, or 0th order, domain wall intersection (109, 43) corresponds to a four-dimensional scale of 5.12 MeV, and lies exactly half-way between the positions corresponding to the masses of the up and down quarks of Family 1 of the Standard Model. The position of the principal intersection (147, 58) corresponds to a scale of 0.18 eV, the quasi-degenerate neutrino mass scale [10]. The position of the 1st order domain wall intersection (14, 5.5) corresponds to a scale of $2 \times 10^{16}$ GeV, the GUT scale of the MSSM. The 1st order intersection (90, 35.5) lies exactly half-way between the positions corresponding to the masses of the bottom and top quarks of Standard Model Family 3. The position of the 2nd order domain wall intersection (99.5, 39.25) corresponds to a scale of 375 MeV, the QCD scale $\overline{\Lambda}$ [11]. This intersection also lies exactly half-way between the positions corresponding to the masses of the strange and charm quarks of Standard Model Family 2. The Rydberg constant $R_\infty$ (13.6 eV) and the Higgs field vacuum expectation value (246 GeV) correspond closely with 2nd and 3rd order domain wall intersections, respectively. The positions of those intersections corresponding to the scales of the Standard Model are shown in Figure 2.

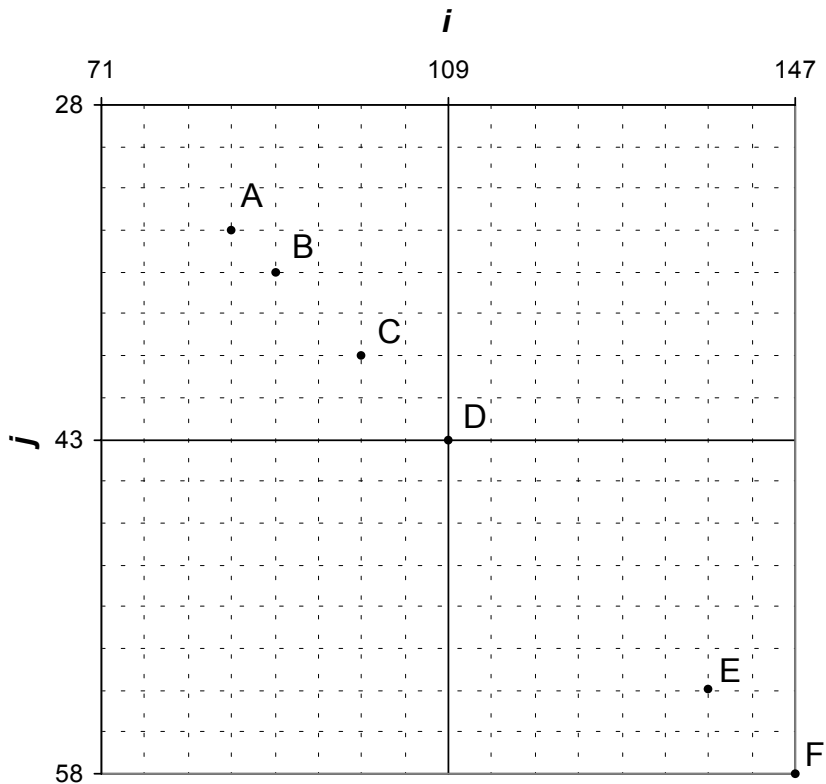

Figure 2: Domain walls in the noncompact extra space of the model, and the positions corresponding to the scales of the Standard Model.

A  Higgs field VEV  
B  Family 3 quark mass scale  
C  Family 2 quark mass scale  
D  Family 1 quark mass scale  
E  Rydberg constant $R_\infty$  
F  Quasi-degenerate neutrino mass scale

# 3  Standard Model scales from the geometry of spacetime

We will hypothesize that the noncompact extra space of the model is the cover of a compact space that is either a parallelogram or a rectangle with sides of length $R\ln(\pi/2)$ and $R\ln\pi$, where $R$ is the AdS radius of curvature. We will denote by (1, 0) and (0, 1) the fundamental 1-cycles of the compact space. The 1-cycle along the diagonal will be denoted by (1, 1). The distance in the covering space from the Planck region (0, 0) to the domain wall intersection $(i, j)$ equals the length of $i$ (1, 0) cycles and $j$ (0, 1) cycles of the compact space. In a rectangular compact space, the (1, 1) cycle will be of length $L = R\{[\ln(\pi/2)]^2 + [\ln\pi]^2\}^{1/2}$. At the position of the principal domain wall intersection (109, 43), which corresponds to the quark mass scale of Standard Model Family 1, we find that

$$109 R \ln(\pi/2) \approx 43 R \ln\pi \approx 40 L. \qquad (3)$$

More precisely,

$$109.0000 R \ln(\pi/2) = 42.9992 R \ln\pi = 39.9994 L. \qquad (4)$$

The principal intersection (109, 43) therefore lies at a distance from the Planck region (0, 0) that is equal to the length of 109 (1, 0) cycles, 43 (0, 1) cycles and 40 (1, 1) cycles of a rectangular compact space (the principal rectangle) with sides of length $R\ln(\pi/2)$ and $R\ln\pi$. And since

$$38 R \ln(\pi/2) \approx 15 R \ln\pi \approx 14 L, \qquad (5)$$

or more precisely

$$38.000 R \ln(\pi/2) = 14.991 R \ln\pi = 13.945 L, \qquad (6)$$

the principal intersection (147, 58), upon which the natural scale is that of the quasi-degenerate neutrino mass scale, lies at a distance from the Planck region (0, 0) that is equal to the length of 147 (1, 0) cycles, 58 (0, 1) cycles and, approximately, 54 (1, 1) cycles of the principal rectangle. Also, the principal intersection (71, 28) lies at a distance from the Planck region that is equal to the length of 71 (1, 0) cycles, 28 (0, 1) cycles and, approximately, 26 (1, 1) cycles of the principal rectangle. 1-cycles winding around the (1, 0) and (0, 1) cycles of the principal rectangle therefore meet, close to the line $y = z$ in the covering space, in the principal domain wall intersections identified earlier.

The distance from the Planck region (0, 0) to the 1st order domain wall intersection (90, 35.5), the position of which corresponds to the quark mass scale of Standard Model Family 3, equals the length of 90 (1, 0) cycles, 35.5 (0, 1) cycles and 33 (1, 1) cycles of the principal rectangle. The fractional number of (0, 1) cycles of the principal rectangle, up to the point of the intersection, can be interpreted as an integer number of (0, 1) cycles of the 1st order rectangle, with sides of length ½$R\ln(\pi/2)$ and ½$R\ln\pi$, which is also covered by the noncompact two-dimensional extra space. The distance from (0, 0) to the 2nd order intersection (99.5, 39.25), the position of which corresponds to the quark mass scale of Standard Model Family 2, equals the length of 99.5 (1, 0) cycles, 39.25 (0, 1) cycles and 36.5 (1, 1) cycles of the principal rectangle. The fractional numbers of (0, 1) and (1, 1) cycles, up to the point of the intersection, can be interpreted as integer numbers of (0, 1) and (1, 1) cycles of the 2nd order rectangle, with sides of length ¼$R\ln(\pi/2)$ and ¼$R\ln\pi$, which is also covered by the noncompact two-dimensional extra space. Similarly, we can interpret 3rd order intersections as the intersections, in the covering space, of 1-cycles winding around the 3rd order rectangle.

1-cycles winding around the (1, 0) and (0, 1) cycles of the principal rectangle and a hierarchy of rectangular subspaces meet, upon the line y = z in the covering space, in intersections upon which the natural scales are the scales of the Standard Model.

## 4  The intersecting brane world

We have shown that the three families of quarks are associated with the intersections (109, 43), (99.5, 39.25) and (90, 35.5). The hadrons and the weak gauge bosons are also associated with domain wall intersections in the noncompact extra space, and therefore with the orthogonal intersections of 1-cycles winding around the compact spaces of the model. 4-branes wrapping the 1-cycles intersect orthogonally in 3-branes, upon which the particles propagate. The neutral hadrons $\pi^0$, $K^0$, $\phi$ and $\Lambda_b^0$, and the exotic baryon $\theta^+$, are located adjacent to the intersections (101.75, 40.125), (98.875, 39), (97.25, 38.375), (93.5, 36.875) and (96.375, 38), respectively, as shown in Figure 3. Each of these hadrons appears to lie precisely upon a low order domain wall in one direction, and upon a high order domain wall in the other direction. Other hadrons are associated with the intersections of domain walls of higher order. For example, $K^{*0}$ is located adjacent to the intersection of domain walls of 4th and 1st order.

Groups of particles are also associated with the intersections of low order domain walls in the noncompact extra space. The groups $\pi^o$ - $\pi^\pm$, $K^\pm$ - $K^0$, $\rho$ - $\omega$, $\Lambda$ - $\Sigma^0$ and $W^\pm$ - $Z^0$ are centred on locations adjacent to the intersections (101.75, 40.125), (98.875, 39), (97.875, 38.625), (97, 38.25) and (87.5, 34.5), respectively, as shown in Figure 4.

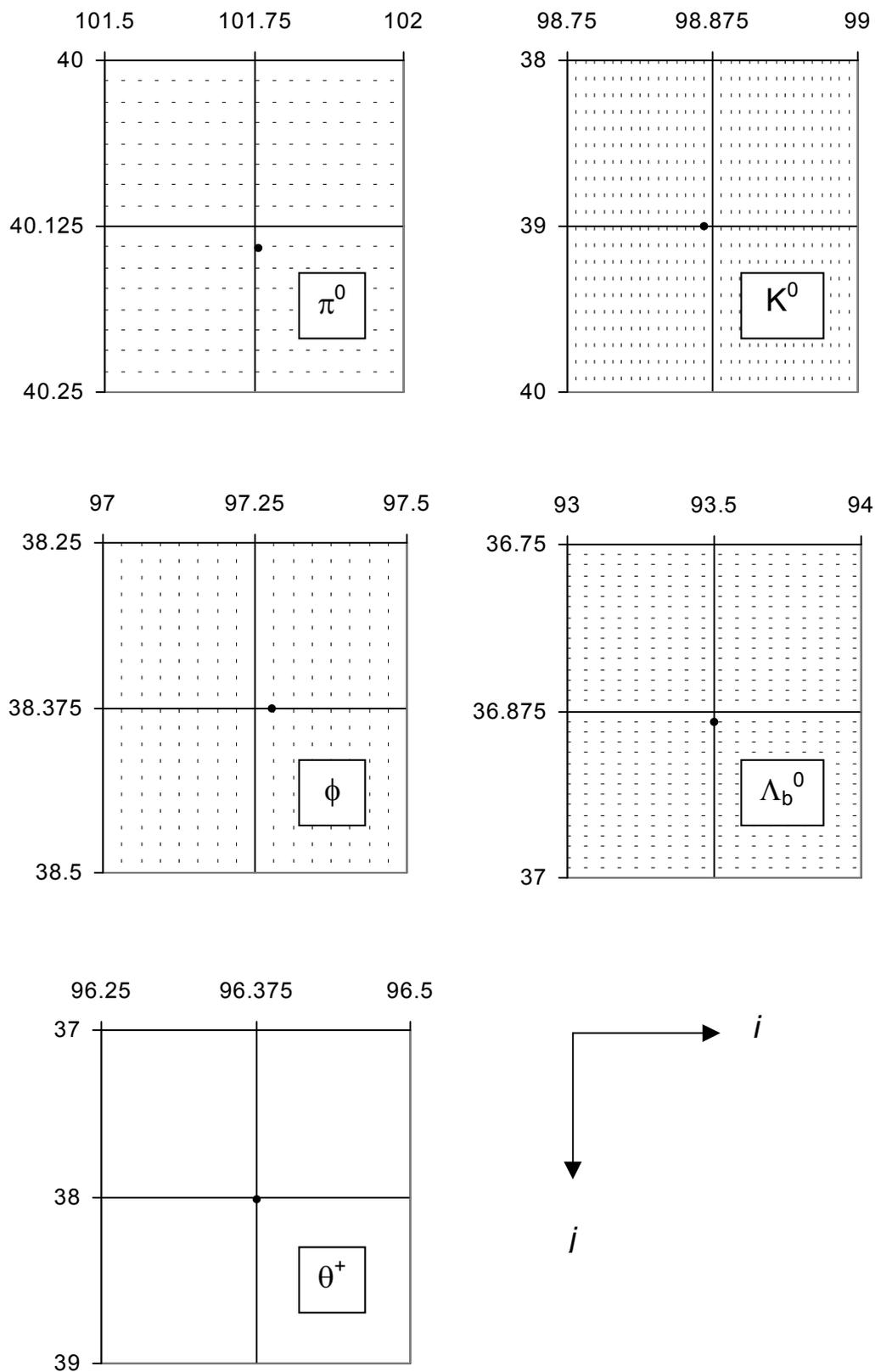

Figure 3: The locations of hadrons upon domain wall intersections in the noncompact extra space of the model.

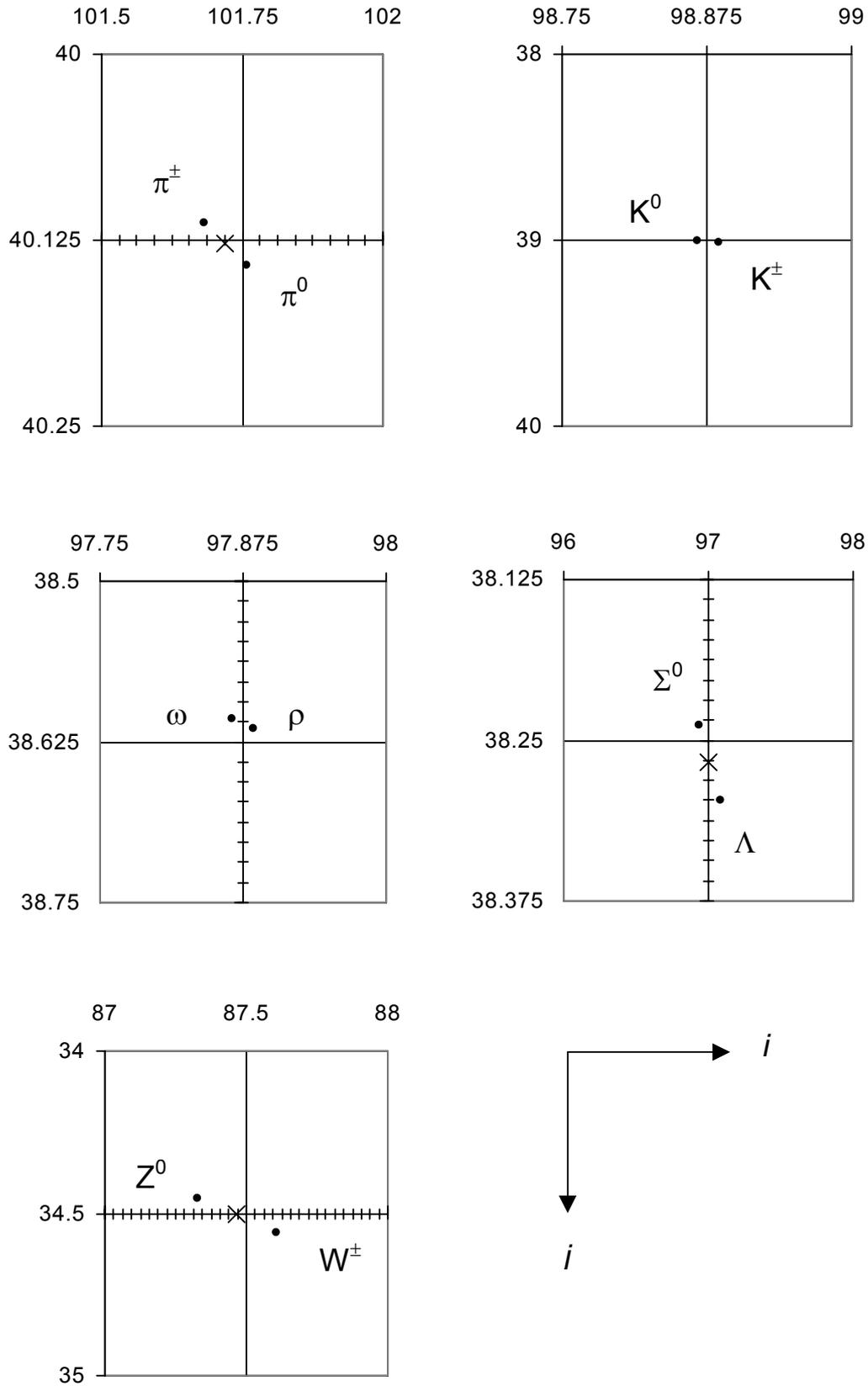

Figure 4: The locations of particle groups in the noncompact extra space of the model. For some groups, a cross marks the geometric mean of the two particle masses.